\documentclass[reprint,aps,prb]{revtex4-2}

\usepackage[pdftex]{graphicx}
\usepackage{amsmath}
\usepackage{comment}

\begin{document}

\title{Chirality in BaTiOCu$_4$(PO$_4$)$_4$}

\author{Alex Hallett}
\affiliation{Materials Department, ETH Zurich}

\author{Nicola A. Spaldin}
\email[Corresponding author: ]{ahallett@ethz.ch}
\affiliation{Materials Department, ETH Zurich}

\date{\today}

\begin{abstract}
We present a first principles study of the ferrochiral phase transition in chiral Ba(TiO)Cu$_4$(PO\textsubscript{4})\textsubscript{4}, which has been shown using X-ray diffraction to proceed via the antiferroaxial rotation of antipolar structural units. We analyze the atomic-site electric and magnetic multipole moments to identify connections between these multipoles and chirality and corroborate the previous experimental interpretation. We show that antiferroically ordered atomic-site electric toroidal dipole moments act as an order parameter for the antiferroaxial rotations, and that the overall chiral order is parameterized by the composite order of the antipolar electric dipole and electric toroidal dipole moments. Finally, we evaluate the suitability of various proposed order parameters for chirality, show that some can be excluded and suggest the most promising directions for future exploration. 
\end{abstract}

\maketitle

\section{Introduction}

Chiral systems, which are non-super-imposable onto their mirror images,  play a key role in many fields including biology and medicine \cite{nguyen_chiral_2006},  and studies of the origin of life \cite{salam_role_1991}. In addition, chiral optical activity is used in sensors and displays \cite{zor_chiral_2019},  and phenomena such as chiral spin selectivity and topologies show promise for use in highly efficient nanoelectronic devices \cite{bloom_chiral_2024}. In spite of the fundamental interest and applications, however, there is not yet a rigorous way of quantifying chirality. 

Whether a crystal is chiral or not can be determined based on symmetry: if a structure has no improper rotation symmetry elements (inversion centers, roto-inversions, or mirrors) which convert right-handed coordinate systems to left-handed ones, then the structure is chiral. With this definition, one can unambiguously categorize the 230 crystallographic space groups as chiral or achiral \cite{fecher_chirality_2022-2}. Within the 65 chiral space groups, called Shonke groups, there is a subgroup of 11 enantiomorphic pairs containing handed screw axes, where the enantiomers are in different space groups determined by the screw axis pair. For example, K$_3$NiO$_2$ is a handed chiral material, with enantiomers of space group P$4_12_1$2 and P$4_32_1$2, the key symmetry elements being the $4_1$ ($4_3$) screw axes with opposite handedness, representing 90$^\circ$ rotations coupled with translations of 1/4 (3/4) the length of the unit cell. 

\begin{figure*}[t]
\includegraphics[width=1\textwidth]{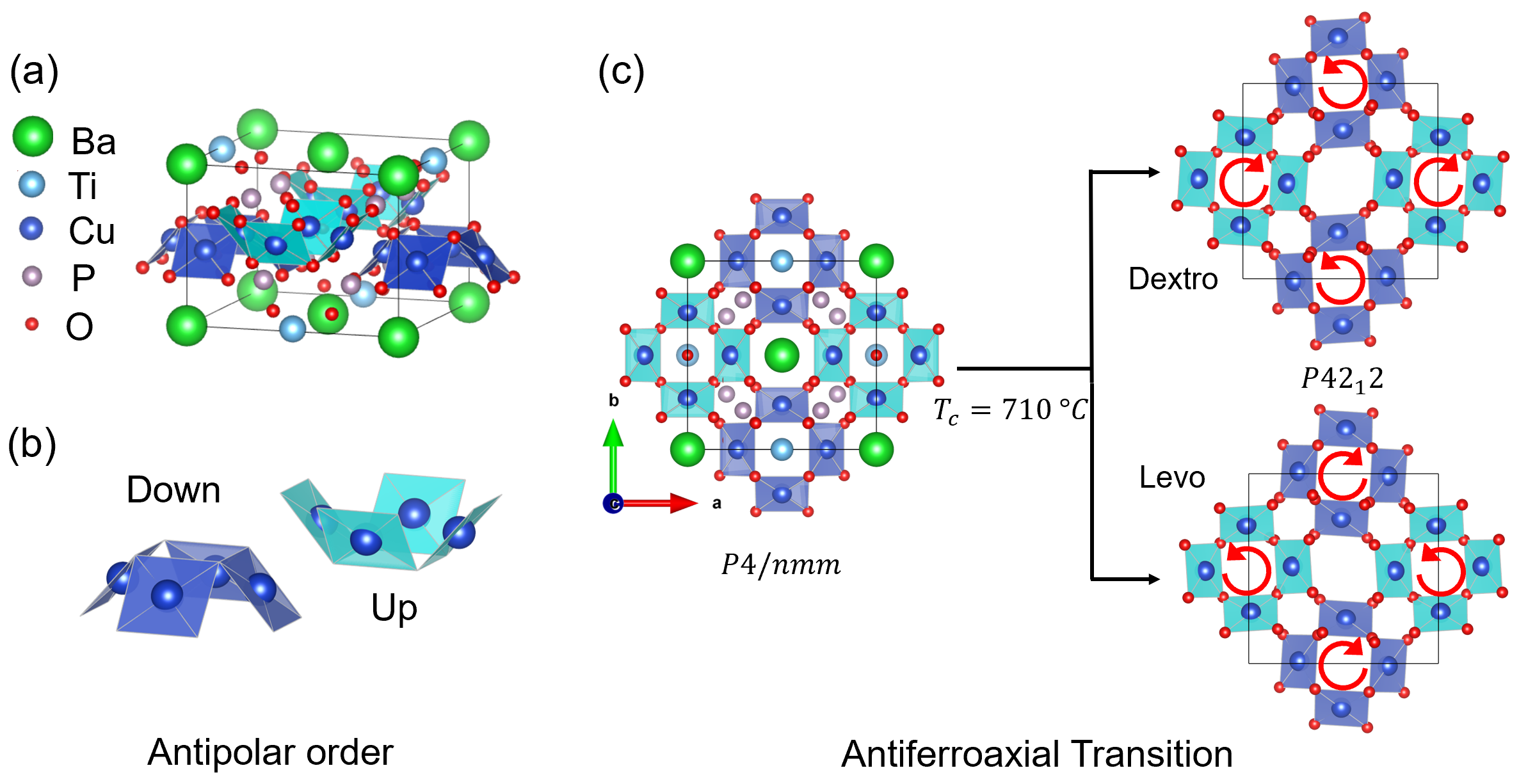}
\caption{Ferrochiral order as a combination of antipolar and antiferroaxial orders. (a) Standard view the high-symmetry crystal structure. (b) Schematic of the up and down facing cupolas, which embed antipolar, local inversion-symmetry breaking order into the high-temperature achiral structure. (c) View along the c axis around which the antiferroaxial rotations occur, breaking the mirror symmetries of the high temperature ($P4/nmm$) structure leading to two chiral enantiomers ($P42_12$) at 710$^\circ$C . Enantiomers correspond to opposite rotation directions of the cupolas.}
 \label{fig:Fig_1_crystal_structure}
\end{figure*}

In this work, we use BaTiCu$_4$(PO$_4$)$_4$ (BTCPO) as a model material to evaluate and compare several proposed quantitative measures of chirality. The defining feature of the BTCPO crystal structure is its antipolar arrangement of polar Cu$_4$O$_{12}$ units, called cupolas, with up and down configurations. The cupolas are comprised of four Cu atoms coordinated by oxygen square planes, which corner share at the base of the cupola. The crystal structure is shown in Fig.~\ref{fig:Fig_1_crystal_structure}(a,b). 

BTCPO undergoes a ferrochiral transition from the high-temperature achiral $P4/nmm$ structure to the chiral P42$_1$2 structure \cite{hayashida_observation_2021} at 710$^\circ$C.  It should be noted that both chiral enantiomers have the same space group symmetry \cite{hayashida_observation_2021,bousquet_structural_2025}, making BTCPO a chiral, but non-handed material. The chiral transition is characterized by antiferroaxial rotations of the cupolas,  where ferroaxiality is generally defined as a spontaneous rotational structural distortion with a resulting axial vector symmetry. The structural transition is shown schematically in Fig.~\ref{fig:Fig_1_crystal_structure}(c), with two opposite rotation patterns resulting in $levo$ and $dextro$ enantiomers depending on the direction of the rotation.

We begin this work by providing first-principles confirmation of the earlier experimental picture that the high-temperature antipolar order, combined with the antiferroaxial ordering across the phase transition lead to the emergence of a chiral phase \cite{hayashida_observation_2021}. We use density functional theory (DFT) to analyze the high and low symmetry phases of BTCPO, as well as the pathway between them. We evaluate the electric toroidal dipole as a measure of the antiferroaxial order, as well as various measures of the electric toroidal monopole which has been proposed as a pseudoscalar order parameter for chirality \cite{oiwa_rotation_2022,kusunose_emergence_2024}. Finally, we analyze several structural quantification parameters for chirality suggested in the literature to determine their suitability for BTCPO.

The remainder of the paper is organized as follows. In Section II we describe our computational methods and present our material characterization via phonon and density of states calculations. In Section IIIa we compare electronic measures for chirality in BTCPO, and in Section IIIb we compare structural measures of chirality. We summarize our conclusions in Section IV. 

\section{Computational Procedure}

Our calculations were performed using density functional theory (DFT) as implemented in the Vienna \textit{ab intio} simulation package (\textsc{vasp})~\cite{kresse_ab_1993, kresse_efficient_1996} using the supplied projector-augmented wave (PAW) potentials~\cite{kresse_ultrasoft_1999} within the generalized gradient approximation (GGA) and Perdew-Burke-Ernzerhof (PBEsol) scheme for solids  ~\cite{perdew_generalized_1996,PBEsol_2008}.  The valence electrons were included for specific elements  as follows: Ba(5s$^2$5p$^6$6s$^2$), Cu(3d$^{10}$ 4p$^{1}$), Ti(3p$^6$4s$^2$3d$^2$), P(2s$^2$2p$^3$),  O(2s$^2$2p$^4$). Electronic wave functions were expanded in a plane-wave basis set with a kinetic energy cutoff of 680~eV. The reciprocal space was sampled using a $2 \times 2 \times 3$ $\Gamma$-centered $k$-point mesh for the 54-atom primitive unit cell the full  list of input parameters is given in the Supplemental \cite{Supplemental_Materials}. 

Starting from the experimental chiral and achiral structures, we relaxed the lattice parameters and atomic positions while constraining the symmetry, including spin polarization at the collinear level, without spin-orbit coupling. The calculated lattice parameters and atomic positions for the achiral and chiral structures are compared to the experimental values from \cite{hayashida_observation_2021} in the Supplemental Material, along with a complete description of all input parameters \cite{Supplemental_Materials}.  

The phonon frequencies of the achiral BTCPO structure at the $\Gamma$-point were calculated using the VASP density functional perturbation theory (DFPT) method with spin polarization excluded.  Since there is no unit-cell change across the chiral transition,  the $\Gamma$-point phonon calculation allowed us identify the imaginary phonon eigenvector corresponding to the chiral instability. 

For density of states and multipole analysis, we use the non-collinear version of VASP, including spin-orbit coupling, with symmetry turned off. To obtain the experimentally observed antiferromagnetic ground state for BTCPO we use the DFT$+U$ method with a Hubbard parameter of 4 eV on the copper atoms. 

Multipole calculations are performed by decomposing the density matrix $\rho_{lm,l'm'}$, obtained from the self-consistent calculations, into irreducible (IR) spherical tensor components $w^{k p r\nu}_t$ \cite{cricchio_itinerant_2009, bultmark_multipole_2009}. Here, the indices $k$, $p$, $r$, $\nu$ and $t$ denote the spatial index, spin index, rank of the tensor, presence or absence of time reversal symmetry and the component of the tensor, respectively. Specifically, $p$ takes values of 0 or 1 for charge and magnetic multipoles, respectively. Time-reversal even and odd symmetries are represented by $\nu=0$ and $\nu=1$, respectively. The rank $r$ ranges from ${ | k - p |}$ to ${ k + p }$, and $t$ ranges from ${ -r}$ to ${ r }$.  We investigate in particular the atomic site electric dipoles and electric toroidal dipoles and their order. The electric dipole is represented by the spherical tensor $w^{1 0 1}_t $ and the electric toroidal dipole is represented by  $w^{1 1 1}_t $ \cite{bhowal_electric_2024}. 

\section{Results and Discussion}

\subsection{Verification of Materials Properties} 

Experimental XRD analysis shows that the achiral to chiral phase transition is characterized by  the antiferroaxial rotation of the oxygens in the cupolas \cite{hayashida_observation_2021}. We confirm this computationally by analyzing the phonon eigenvectors of our relaxed structures. For the achiral structure, we identify a single, unstable, imaginary phonon mode at $\Gamma$ with a frequency of -0.47i THz. The eigenvector of this mode corresponds closely to the displacements connecting the achiral to chiral experimental structures. Spin polarization is not needed to reproduce the chiral instability. The full list of input parameters for the phonon calculation, as well as a plot of the eigenvector are available in \cite{Supplemental_Materials}.

The $3d^9$ 1$\mu B$ Cu${2+}$ spin magnetic dipole moments order antiferromagnetically at 9.5 K in BTCPO \cite{rasta_magnetic_2020,kimura_magnetodielectric_2016}. The spins are oriented perpendicular to the square-planar Cu-O planes within the cupolas, and are arranged antiferroically. The calculated and experimental magnetic ground state are shown in  \cite{Supplemental_Materials}. 

Magnetic reflections measured by neutron diffraction indicate a single propagation wavevector $k=(0,0,0.5)$ indicating an additional unit cell doubling along the [001] direction \cite{kimura_magnetodielectric_2016,kimura_-cation_2018}. For the multipole calculations discussed in Section III(A) we use a single, antiferromagnetic unit cell, neglecting the doubling along the c axis, after verifying that the magnetic $k=(0,0,0.5)$ wavevector has no effect on the electric multipoles relevant to this work \cite{Supplemental_Materials}. 
\begin{figure}[ht]
\centering
\includegraphics[width=1\columnwidth]{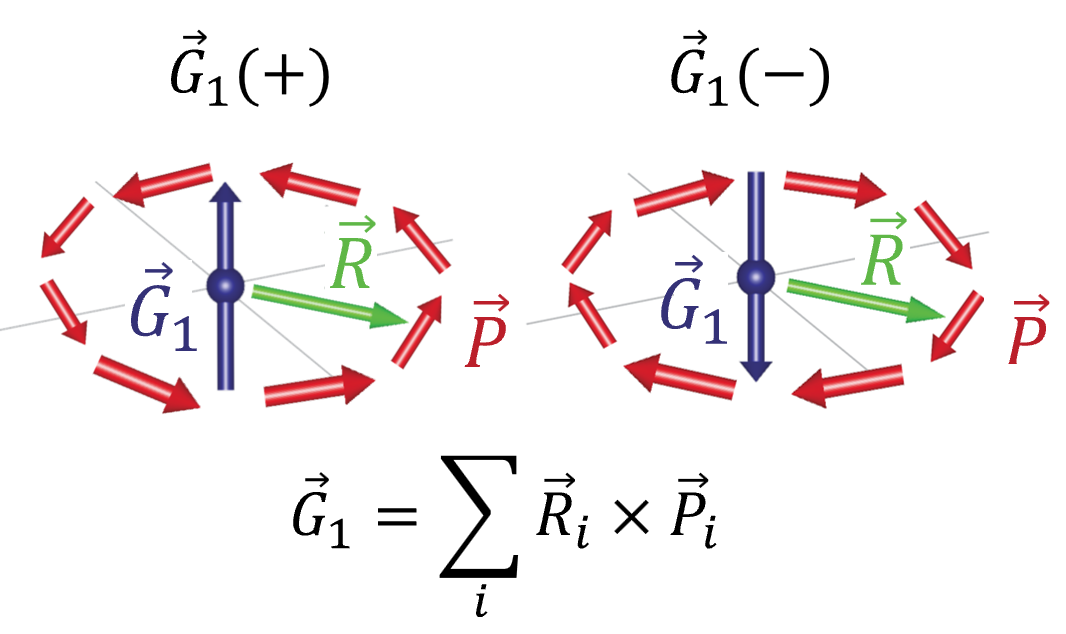}
\centering
\caption{\textbf{Electric Toroidal Dipole} 
Schematic of electric toroidal dipole moments (blue arrows) resulting from rotational patterns of electric dipoles (red arrows). Oppositely handed electric dipoles have opposite electric toroidal dipoles.}
\label{fig:Fig_electric_toroidal_dipole}
\end{figure} 

The antiferroaxial phase transition is geometric, and the structural distortion is not accompanied by a substantial change in hybridization of the atomic orbitals, which is expected since rotational distortions typically do not result in changes in the orbital character \cite{hill_why_2000}. Our density of states calculations reflect this pattern indicating there is no significant change in orbital hybridization and the transition can be described as geometric \cite{Supplemental_Materials}. 

\subsection{Electronic Quantification of Chirality} 
\begin{figure*}[ht!]
\centering
\includegraphics[width=1\textwidth]{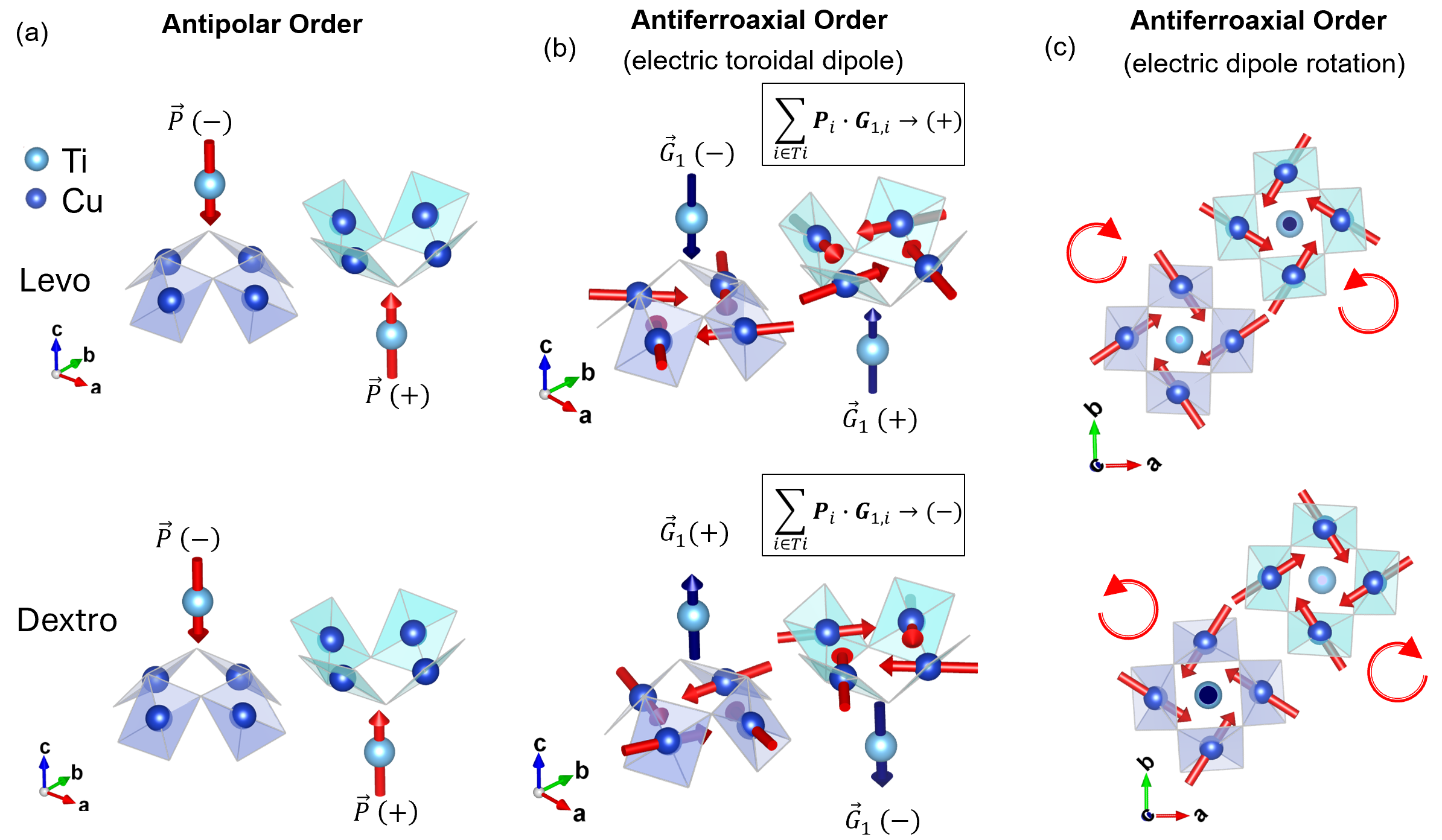}
\centering
\caption{\textbf{Multipole moments representing antipolar and antiferroaxial order.} 
(a) Up- and down-oriented cupolas, with electric dipole moments, calculated as the $w^{101}$ spherical tensor components from the electric multipole expansion, shown as red arrows on the out-of-plane Ti atom in the center of the cupola.
The dipole points in the direction of the opening of the cupola. The antipolar order is set by the direction of the cupola is unchanged by the antiferroaxial rotation patterns, so is the same for both enantiomers. Domains of opposite antipolar order are  not considered here. (b) The rotating dipole moments in the $xy$ plane generate out-of-plane electric toroidal dipole $w^{111}$ moments (blue) on the Ti atoms.
Opposite rotations of the dipole moments create oppositely oriented electric toroidal dipole moments. The scalar of the dipole moments $P$ and the electric toroidal dipole moments $G_1$ yield a ferroically aligned parameter with the same symmetry as the electric toroidal monopole (time-reversal even, parity odd). (c) The electric dipole $w^{101}$ moments (red) form a rotational pattern looking down along the c-axis. 
In each enantiomer, the pattern of dipole moments and the corresponding orientation of the electric toroidal dipoles are reversed. 
}
\label{fig:Fig_multipole_moments}
\end{figure*} 
\begin{figure*}[ht!]
\centering
\includegraphics[width=1\textwidth]{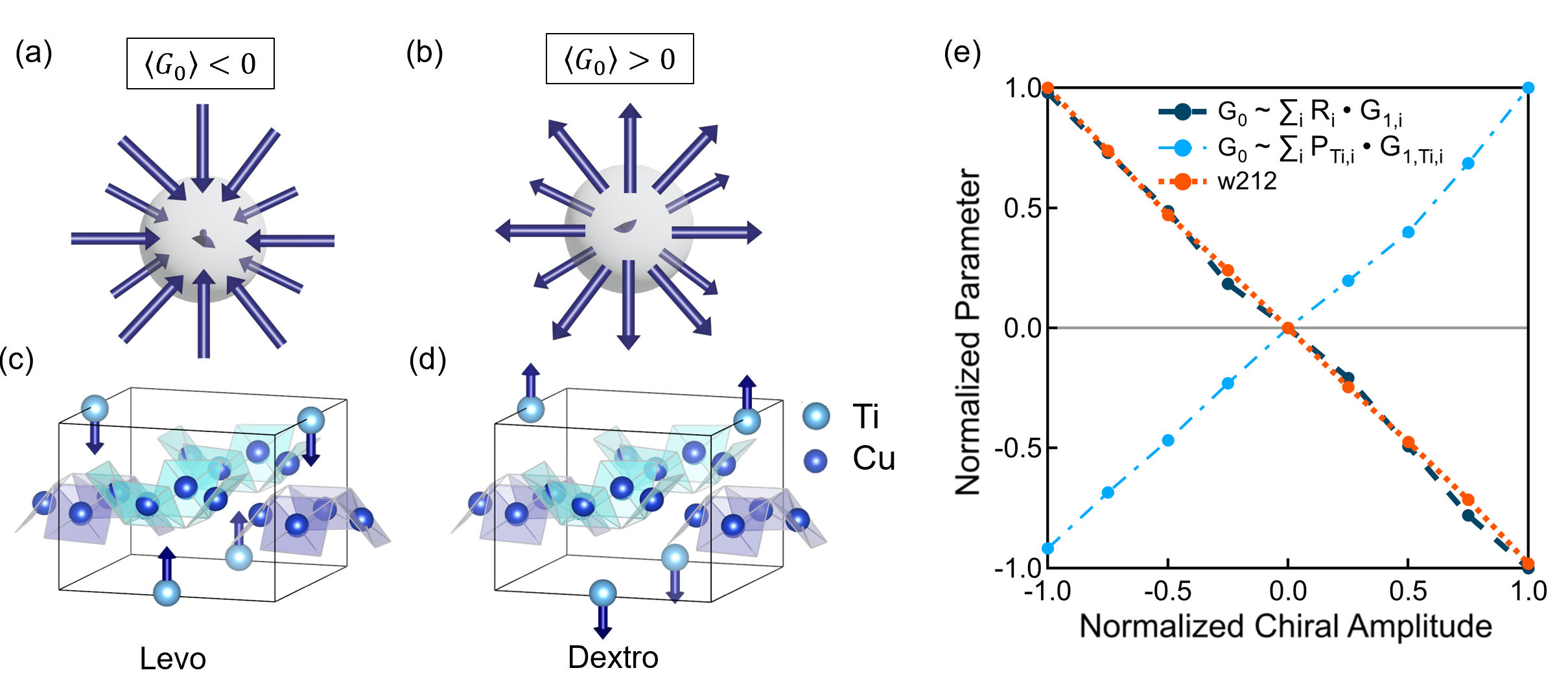}
\caption{\textbf{Electric toroidal monopole versus chiral amplitude.} (a) The negative $G_0$ pattern is shown schematically as a sink of $G_1$ moments, shown as blue arrows. (b) The positive $G_0$ pattern is shown schematically a source of $G_1$ moments, shown as blue arrows. (c) The Levo enantiomer displays an inward-pointing pattern of $G_1$ moments on the Ti atoms.  (d) The Dextro enantiomer displays an analogous outward pointing pattern of $G_1$ moments on the Ti atoms. (e) The normalized electric toroidal monopole $\tilde{G}_0$, calculated from the dot product definition in Eq. \ref{eq:site_G0} is plotted as a dark blue dashed line versus the amplitude of the chiral distortion. The light blue dashed-dotted line is the product of the electric toroidal dipole and electric dipole moment on the Ti atoms in the cupolas, and the orange dashed line is the sum of the $w^{212}$ moments over all atoms. Here, normalized chiral amplitude ($NCA$) $NCA=-1$ represents the Levo enantiomer, zero corresponds to the achiral reference structure and $+1$ to the Dextro enantiomer.  }
\centering
\label{fig:Fig_G0}
\end{figure*} 

We extract both the local atomic electric dipole ($\textbf{P}$) and electric toroidal dipole ($\textbf{G}_1$) moments from the DFT-calculated density matrix to confirm that the picture obtained from inspecting the structure is reflected at the electronic level.  The electric toroidal dipole (ETD), shown schematically in Fig.~\ref{fig:Fig_electric_toroidal_dipole}, is an axial vector invariant under both time reversal ($\mathcal{T}$) and inversion symmetry ($\mathcal{I}$). It  has been proposed as the order parameter for ferroaxial materials, particularly in cases where rotational dipole arrangements of dipoles cause the axial ordering \cite{bhowal_electric_2024}. An atomic scale description of $\textbf{G}_1$ can also be defined from two $\mathcal{T}$-odd vectors $\textbf{G}_1 \propto \textbf{l} \times \textbf{s}$  where $\vec{l}$  and $\vec{s}$  are the orbital and spin angular momenta, respectively.  The quantities $\textbf{l} \times \textbf{s}$ and $\sum \textbf{R} \times \textbf{P}$ have the same symmetry under time reversal and space inversion and are permitted within the same space groups. We extract the atomic-site ETD from the spherical tensor $w^{1 1 1}_t \propto (\textbf{l} \times \textbf{s})$ \cite{laan_core_1995}, with ($l_1+l_2=$ even), since the ETD is even under inversion.  

In Fig.~\ref{fig:Fig_multipole_moments}(a) we show the rotational pattern of electric dipole moments (red arrows) obtained from the $w^{101}$ spherical tensor components of the decomposed density matrix on the copper atoms within the cupola. The electric dipoles on the Cu atoms can have either counterclockwise or clockwise rotations for both the up and down facing cupolas while the orientation of the electric dipole on the Ti atom is fixed by the orientation of the cupola, so that in principle four combinations of polar and ferroaxial dipole orientations for each individual cupola, ($+$, $+$) ($-$,$-$) ($+$,$-$), ($-$,$+$) are possible. We find, however, that the product of these orders, quantified by the scalar product of the electric dipole and the electric toroidal dipole on the titanium atoms ($\boldsymbol{P} \cdot \boldsymbol{G}_1$) always has the same sign on the two cupolas in a given enantiomer . That is, the ($\boldsymbol{P} \cdot \boldsymbol{G}_1$)  composite is ferroically ordered. Therefore, the experimental observation that the chiral order in BTCPO is a composite of macroscopic antipolar and antiferroaxial ordering \cite{hayashida_observation_2021} is also reflected at the atomic level through the composite multipole of  ($\boldsymbol{P} \cdot \boldsymbol{G}_1$). 

The scalar product ($\boldsymbol{P} \cdot \boldsymbol{G}_1$) has the same time reversal and parity symmetry as the electric toroidal monopole (ETM) \cite{Fava_2025}, which is the zeroth order term in the electric toroidal multipole expansion, and has been proposed as an order parameter for crystal chirality \cite{kusunose_emergence_2024,kusunose_symmetry-adapted_2023}. The ETM has been investigated as a potential microscopic order parameter for chirality because it is only allowed by symmetry in the point groups which correspond to chiral crystals. The ETM is formally given by the divergence of the $\textbf{G}_1$ field; 
\begin{equation}
G_0 = \int_V \nabla \cdot \textbf{G}_1(r)  \space d^3r
\label{eqn:G0_int}
\end{equation}
Cartoons showing arrangements of local $\textbf{G}_1$moments giving rise to non-zero and opposite divergences of $\textbf{G}_1$ are shown in Fig.~\ref{fig:Fig_G0}(a,b). In Fig.~\ref{fig:Fig_G0}(c,d), we show the atomic-site $\textbf{G}_1$ moments in the BTCPO enantiomers of BTCPO and see that similarly to the ``hedgehogs" in Fig.~\ref{fig:Fig_G0}(a,b) all $G_1$ moments point either inwards, or outwards away from the center of the unit cell. Capturing these observations, an alternative formulation of the integral in Eq. \ref{eqn:G0_int} has been presented, which has the same symmetry as the formal monopole, but is more tractable to calculate: 
\begin{align}
\tilde{G}_0 &= \sum_n \ \mathbf{R}_n \cdot \mathbf{G}_1(\mathbf{R}_n)  \quad .
\label{eq:site_G0}
\end{align}
We evaluate the quantity in Eq. \ref{eq:site_G0} across the chiral phase transition, using the center of each cupola as an origin point, contributions of each cupola are summed together. In Fig.~\ref{fig:Fig_G0}(e), the electric toroidal monopole $\tilde{G}_0$ as calculated from Eq. \ref{eq:site_G0} is shown as a dark blue dotted line as a function of the amplitude of the chiral distortion.  The magnitude of $\tilde{G}_0$ decreases linearly from one enantiomer to another, passing through zero when the structure is achiral. Details on the choice of reference and the origin are shown in the \cite{Supplemental_Materials}. Note that the electric toroidal monopole in this case acts as a sign-sensitive measure of chirality across the phase transition. 

Finally, we search among the higher order electric toroidal multipoles up to spatial index $k=2$ for possible chiral order parameters. Only one moment --- the $t=0$ component of the $w^{212}$ moment with $\nu = 0 $, $p=1$--- showed suitable behavior, being zero in the achiral phase and having opposite sign in the two enantiomers. This moment is time reversal even, inversion odd, and has $l_1 +l_2 =$ odd, as required for a chiral order parameter.  This multipole moment is proportional to a component of the second order quadrupole moment ($G_2$) in the electric toroidal expansion. The atomic-site $w^{212}$ moments are ordered ferroically within each atomic species, with the Ba moments having opposite orientation from the Ti and P atoms. The normalized sum of the atomic contributions for all species is plotted as the orange dashed line in Fig.~\ref{fig:Fig_G0}(e). The $w^{212}$ moment has the same behavior as the $G_0$ parameter across the transition. A description of the mathematics of $G_2$ and the higher order electric toroidal dipole moments is an intriguing direction for future work.  

\subsection{Structural Quantification of Chirality}

\begin{figure}[ht!]
\centering
\includegraphics[width=\columnwidth]{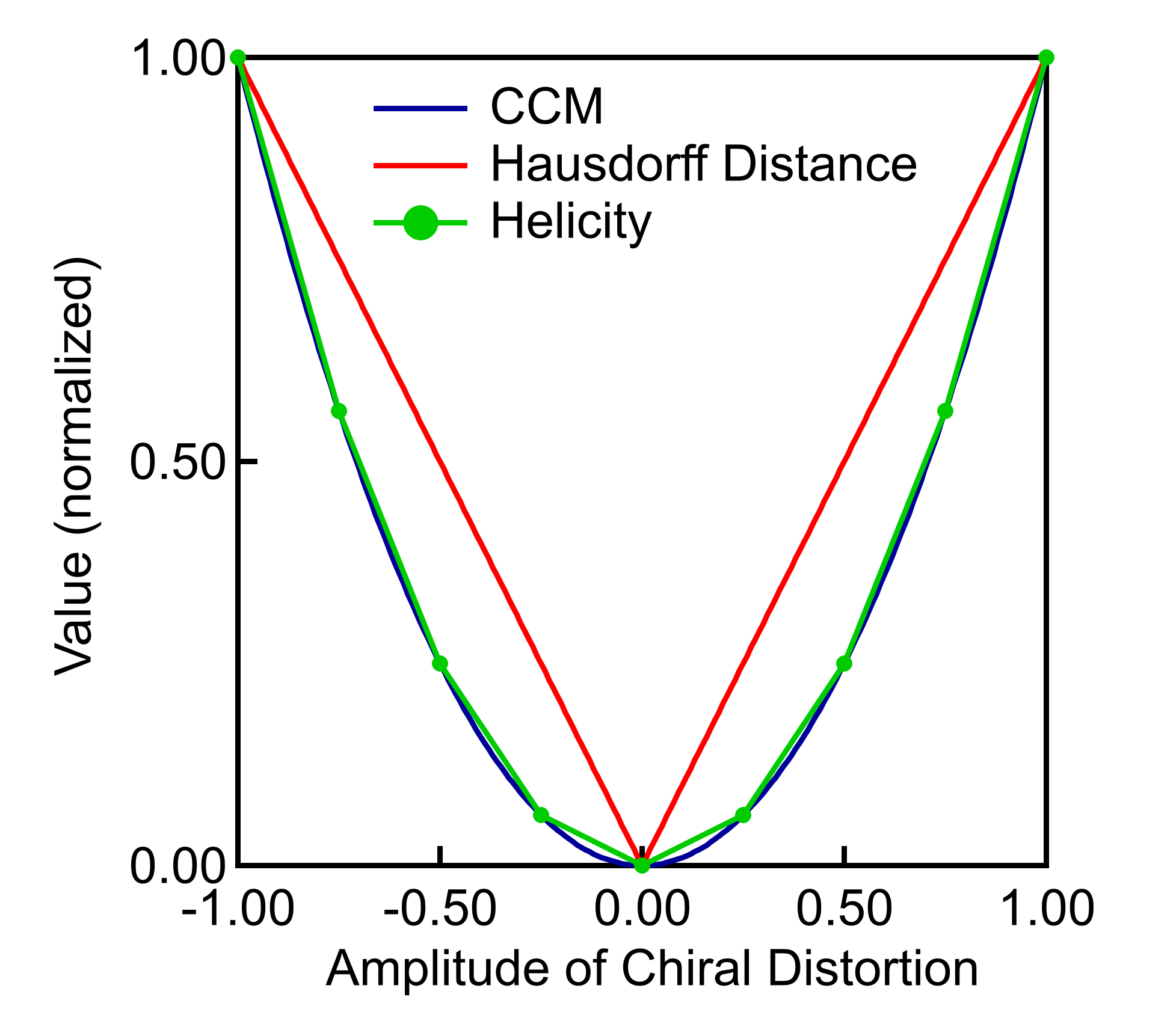}
\centering
\caption{\textbf{Structural Measures of Chirality} A comparison of different structural parameters for chirality including the continuous chirality measure, Hausdorff distances, and helicity. The CCM and helicity show a parabolic trend across the pathway between enantiomers, while the Hausdorff distance scales linearly from the achiral reference phase to its maximum amplitude at each enantiomer. None of the parameters are sign sensitive between enantiomers, and all necessitate the existence of an achiral reference phase.}
\label{fig:Fig_quantifying_chirality}
\end{figure}

Finally, we analyze the suitability of three structural chirality measures that have been used in the literature --- the continuous chirality measure (CCM), the Hausdorff distance and the helicity --- for the case of BTCPO and show our results in Fig.~\ref{fig:Fig_quantifying_chirality}. The continuous chirality measure quantifies the relationship between the chiral and achiral reference structures based on the mean square of the distances between the atomic positions, expressed mathematically as
\begin{align}
 \mathrm{CCM} = \frac{1}{N} \sum_{i=1}^{N} \lVert \vec{x}_i - \vec{x}'_i \rVert^2 \quad .
 \label{eqn:CCM}
 \end{align}
Here, ${x}$ and $x'$ are the atomic positions in the achiral and chiral structures respectively, $i$ indicates the atomic site, and $N$ is the total number of atoms. The dark blue line in Fig.~\ref{fig:Fig_quantifying_chirality} shows the CCM plotted versus the normalized amplitude of the chiral distortion from one chiral enantiomer, through the achiral reference to the other enantiomer. We see that there is a parabolic dependence on the chiral distortion as expected from Eq. \ref{eqn:CCM}. The CCM is therefore not an adequate order parameter for chirality, as it does not distinguish between opposite enantiomers and also requires an achiral reference structure. 

Next, we consider the Hausdorff distance between chiral and achiral structures. The Hausdorff distance ($H(q)$)  depends on the distances $d(\hat{x}, \hat{x'}) = \sqrt{(x-x')^2}$ between the points $\hat{x}$ of the chiral structure and points $\hat{x}'$ of the reference structure and is defined as 
\begin{align}
H(q) &= \frac{h(q)}{h_{\mathrm{maximum}}}\,, \\[0.8em]
&\text{where}\\[0.8em]
h(q) &= \sup_{\hat{x} \in X}\, g(\hat{x}(q)) \\[0.5em]
&\text{and}\\[0.5em]
g(\hat{x}(q)) &= \inf_{\hat{x}' \in X_{\mathrm{ref}}} d(\hat{x}(q), \hat{x}')\quad.
\end{align}
Here, $\sup$ is the supremum, the least upper bound, while $\inf$ is the infinimum, the greatest lower bound. The normalized Hausdorff distance is plotted as the red line in Fig.~\ref{fig:Fig_quantifying_chirality}, showing a linear dependence on the chiral amplitude. Like the CCM, the Hausdorff distance cannot distinguish between enantiomers and requires the existence of an achiral reference phase for its definition. 

A third parameter, helicity ($\mathcal{H}$) has been recently proposed as a measure of structural chirality \cite{gomez-ortiz_structural_2024}.  The definition is based on helicity of flow from fluid dynamics, where the velocities of the fluid $\mathbf{v}$ are replaced by the trajectories of the atomic positions across the chiral phase transition for arbitrary units of time, 
\begin{align}
\mathcal{H} = \int d^3\mathbf{r} \, \mathbf{v} \cdot [\nabla \times \mathbf{v}]\quad.
 \end{align}
All handed chiral space groups contain handed screw axes with enantiomers in different space groups, and for these handed chiral crystals the helicity would be sign sensitive, having opposite sign for each enantiomer. In Fig.~\ref{fig:Fig_quantifying_chirality}, the helicity is shown in green. We see that its form is the same as the CCM and that it does not distinguish between enantiomers. In the case of BTCPO, both enantiomers are within the same space group, which contains only a neutral screw $2_1$ screw axis, and thus the helicity cannot act as a sign-sensitive order parameter. 

In summary, all three parameters have two factors which make them unsuitable as a generalized chiral order parameter: they do not distinguish between enantiomers for non-handed space groups, and they require the presence, at least in principle, of an achiral reference phase.

\section{Summary}

Through our DFT structural and multipole analysis we confirm computationally the earlier experimental interpretation that the combination of antipolar ($P$) and antiferroaxial ($G_1$) orders in BTCPO leads to chiral order \cite{hayashida_switching_2022}. The antipolar ordering of the cupolas remains constant in orientation and magnitude across the chiral transition, and is captured in our calculations by the out-of-plane local electric dipole moment ($\textbf{P}$) on the Ti atoms in the center of the cupolas. A second, rotational, electric dipole ordering on the Cu atoms is associated with the antiferroaxial phase transition, and leads to out-of-plane electric toroidal dipole moments ($\textbf{G}_1$) on the Ti atoms, which we calculate from the $w^{111}$  multipole moments. 

Both of the electric toroidal monopoles defined as $\sum_i \textbf{R}_i \cdot \textbf{G}_{1,i}$ and the quantity $\textbf{P} \cdot  \textbf{G}_1$ are suitable parameters to quantify chirality. They are both zero in the achiral phase and change sign between BTCPO enantiomers, even though the enantiomers share the same non-handed space group symmetry.  In addition to the electric toroidal monopole, we identify a higher order multipole moment,  $w^{212}$, which displays the same qualitative behavior as $G_0$.  These quantities, ($\textbf{P}\cdot \textbf{G}$), $G_0$, $w^{212}$, act as sign-sensitive order parameters for chirality, in contrast to the proposed structural measures shown in Fig.~\ref{fig:Fig_quantifying_chirality}, which do not distinguish between the enantiomers for this non-handed system. We hope that our work will motivate experimental efforts to probe these multipolar orders in BTCPO and other chiral systems. 

\section*{Acknowledgments}

We thank Andrea Urru, and Carl Romao for helpful discussions. Use was made of computational facilities at the Swiss National Supercomputing center (CSCS) and the Euler cluster. This research was funded by ETH Zurich and the Swiss National Science Foundation (SNSF) under grant number TMPFP2\_234033 . For the purpose of open access, a CC BY public copyright license is applied to any author accepted manuscript (AAM) version arising from this submission. 

\bibliography{library_30_4}

\end{document}